\input epsf

\magnification\magstephalf
\overfullrule 0pt
\def\gsim{\raise.3ex\hbox{$\;>$\kern-.75em\lower1ex\hbox{$\sim$}$\;$}}

\font\rfont=cmr10 at 10 true pt
\def\ref#1{$^{\hbox{\rfont {[#1]}}}$}


\font\fourteenbf=cmbx12 scaled\magstep1

\font\tenbfit=cmbxti10
\font\sevenbfit=cmbxti10 at 7pt
\font\fivebfit=cmbxti10 at 5pt
\newfam\bfitfam 
\textfont\bfitfam=\tenbfit  \scriptfont\bfitfam=\sevenbfit
\scriptscriptfont\bfitfam=\fivebfit

\font\eightit=cmti8

\font\tenbfit=cmbxti10
\font\sevenbfit=cmbxti10 at 7pt
\font\fivebfit=cmbxti10 at 5pt
\newfam\bfitfam 
\textfont\bfitfam=\tenbfit  \scriptfont\bfitfam=\sevenbfit
\scriptscriptfont\bfitfam=\fivebfit

\font\tenbit=cmmib10
\newfam\bitfam
\textfont\bitfam=\tenbit%

\font\tenmbf=cmbx10
\font\sevenmbf=cmbx7
\font\fivembf=cmbx5
\newfam\mbffam
\textfont\mbffam=\tenmbf \scriptfont\mbffam=\sevenmbf
\scriptscriptfont\mbffam=\fivembf

\font\tenbsy=cmbsy10
\newfam\bsyfam 
\textfont\bsyfam=\tenbsy%

   
 \def \f {\phi}

\def\pd {\partial}
\def\pmb#1{\setbox0=\hbox{#1}
 \kern.05em\copy0\kern-\wd0 \kern-.025em\raise.0433em\box0 }

\def\slash{/\kern-.5em}


 %


\def\boxit#1{\vbox{\hrule\hbox{\vrule\kern1pt\vbox
{\kern1pt#1\kern1pt}\kern1pt\vrule}\hrule}}

\def\h{\hfill\break}
\parskip=6pt
\parindent=0pt
\hsize=17truecm\hoffset=-5truemm
\vsize=23truecm
\def\footnoterule{\kern-3pt
\hrule width 17truecm \kern 2.6pt}


\catcode`\@=11 

\def\nolabels{\def\wrlabeL##1{}\def\eqlabeL##1{}\def\reflabeL##1{}}
\def\writelabels{\def\wrlabeL##1{\leavevmode\vadjust{\rlap{\smash%
{\line{{\escapechar=` \hfill\rlap{\sevenrm\hskip.03in\string##1}}}}}}}%
\def\eqlabeL##1{{\escapechar-1\rlap{\sevenrm\hskip.05in\string##1}}}%
\def\reflabeL##1{\noexpand\llap{\noexpand\sevenrm\string\string\string##1}}}
\nolabels
\global\newcount\refno \global\refno=1
\newwrite\rfile
\def\defref{$^{{\hbox{\rfont [\the\refno]}}}$\nref}
\def\nref#1{\xdef#1{\the\refno}\writedef{#1\leftbracket#1}%
\ifnum\refno=1\immediate\openout\rfile=refs.tmp\fi
\global\advance\refno by1\chardef\wfile=\rfile\immediate
\write\rfile{\noexpand\item{#1\ }\reflabeL{#1\hskip.31in}\pctsign}\findarg}
\def\findarg#1#{\begingroup\obeylines\newlinechar=`\^^M\pass@rg}
{\obeylines\gdef\pass@rg#1{\writ@line\relax #1^^M\hbox{}^^M}%
\gdef\writ@line#1^^M{\expandafter\toks0\expandafter{\striprel@x #1}%
\edef\next{\the\toks0}\ifx\next\em@rk\let\next=\endgroup\else\ifx\next\empty%
\else\immediate\write\wfile{\the\toks0}\fi\let\next=\writ@line\fi\next\relax}}
\def\striprel@x#1{} \def\em@rk{\hbox{}} 
\def\lref{\begingroup\obeylines\lr@f}
\def\lr@f#1#2{\gdef#1{\defref#1{#2}}\endgroup\unskip}
\newwrite\lfile
{\escapechar-1\xdef\pctsign{\string\%}\xdef\leftbracket{\string\{}
\xdef\rightbracket{\string\}}}

\def\writestop{\def\writestoppt{\immediate\write\lfile{\string\p
ageno%
\the\pageno\string\startrefs\leftbracket\the\refno\rightbracket%
\string\def\string\secsym\leftbracket\secsym\rightbracket%
\string\secno\the\secno\string\meqno\the\meqno}\immediate\closeout\lfile}}
\def\writestoppt{}\def\writedef#1{}
\catcode`\@=12 
\centerline{\fourteenbf Perturbative QCD and Regge theory: closing the circle}
\vskip 8pt
\centerline{A Donnachie}
\centerline{Department of Physics, Manchester University}
\vskip 5pt
\centerline{P V Landshoff}
\centerline{DAMTP, Cambridge University$^*$}
\footnote{}{$^*$ email addresses: sandy.donnachie@man.ac.uk, \ pvl@damtp.cam.ac.uk}
\bigskip
{\bf Abstract}
We explain how Regge theory and perturbative evolution may be made compatible
at small $x$. The result not only gives striking support to the two-pomeron
description of small-$x$ behaviour, but gives a rather clean test of
perturbative QCD itself. When $x$ is very small, 
the proton's gluon distribution function is significantly larger
than is commonly believed. Perturbative evolution is invalid below
$Q^2\approx 5$~GeV$^2$.

\bigskip\bigskip
{\bf 1 Introduction}

There is a growing realisation\defref\fr{
J Forshaw and D A Ross, {\it Quantum Chromodynamics and the Pomeron},
Cambridge University Press (1997)
}\defref\salam{
M Ciafaloni, D Colferai and G P Salam, JHEP 0007 (2000) 054
}\defref\thorne{
R S Thorne, Physical Review D64 (2001) 074005
}
that the conventional approach  to perturbative evolution breaks down
at small $x$. Certainly, when, at a given fixed small $x$, the structure function
varies rapidly with $Q^2$, it is not
sufficient to expand the DGLAP
splitting function in powers of $\alpha_s$ and truncate the expansion
after one or two terms. Rather, a resummation is needed and, while
there are models of how to do this, so far no reliable method is available.

The Regge approach is apparently orthogonal to that of perturbative evolution.
We have shown\defref\twopom{
A Donnachie and P V Landshoff,  Physics Letters B518 (2001) 63
}
that Regge theory gives a very good description of the data, not only for 
the proton structure function
$F_2(x,Q^2)$, but also for its charm component $F_2^c(x,Q^2)$
and for $J/\psi$ photoproduction. The data indicate rather clearly
the need for a second pomeron, whose trajectory has intercept
$(1+\epsilon_0)\approx 1.4$, 
in addition to the familiar soft pomeron with intercept $(1+\epsilon_1)$ about
1.08. We call the second pomeron the hard pomeron. While at one time
it was hoped that one might calculate its intercept from the BFKL
equation, it now seems likely\defref\forte{
G Altarelli, R D Ball and S Forte, Nuclear Physics B599 (2001) 383
}
that such a calculation is not within
the scope of perturbative QCD. As we explain in this paper, perturbative QCD
merely governs how the magnitude of the hard pomeron's contribution
to the structure function increases with $Q^2$. 

Our belief is\defref\cudell{
J R Cudell, A Donnachie and P V Landshoff, Physics Letters B448 (1999) 281
}\ref{\twopom}
that the hard pomeron is already
present in real-photon amplitudes and contributes with the same 
fixed power $(W^2)^{\epsilon_0}$  at all values of $Q^2$.
That is, at sufficiently large $W$
$$
F_2(x,Q^2)\sim f_0(Q^2) x^{-\epsilon_0}
\eqno(1)
$$
for all values of $Q^2$ and, again  at sufficiently large $W$,
$$
\sigma^{\gamma p}\sim 4\pi^2\alpha_{\hbox{\sevenrm EM}} X_0\,(W^2)^{\epsilon_0}
\eqno(2)
$$
with
$$
X_0=(Q^2)^{-1-\epsilon_0}f_0(Q^2)\Big
\arrowvert_{Q^2=0}
\eqno(3)
$$

There is no theoretical reason why the behaviour should be a power rather
than something more complicated, but we have found\ref{\twopom} that a power 
fits the data very well. If $X_0$ is to be finite,
$f_0(Q^2)$ must vanish as $(Q^2)^{1+\epsilon_0}$ at $Q^2=0$. 
A very good fit to data, all the way from $Q^2=0$ to 5000~GeV$^2$,
is provided by the economical parametrisation
$$
f_0(Q^2)=X_0~(Q^2)^{1+\epsilon_0}/(1+Q^2/Q_0^2)^{1+{1\over 2}\epsilon_0}
\eqno(4)
$$
with $Q_0\approx 3$ GeV.
In this paper, we show that this form agrees remarkably well
with what is obtained from 
DGLAP evolution, over a large range of $Q^2$.  
At present, it is not possible properly to apply perturbative evolution
to the soft-pomeron component of $F_2(x,Q^2)$.
\bigskip

{\bf 2 The DGLAP equation}

\def\u{{\bf u}}\def\P{{\bf P}}\def\f{{\bf f}}
The singlet DGLAP equation is\defref\esw{
R K Ellis, W J Stirling and B R Webber, {\it QCD and Collider Physics},
Cambridge University Press (1996)
}
$$
{\pd\over\pd t}\u (x,Q^2)=\int _x^1 dz\, \P (z,\alpha_s(Q^2))
\,\u({x\over z},Q^2)
\eqno(5)
$$
where $\P$ is the splitting matrix, $t=\log (Q^2/\Lambda^2)$ and
$$
\u= \left (\matrix{x\sum _f(q_f+\bar q_f)\cr xg\cr}\right )
\eqno(6)
$$
Write the Mellin transforms
$$
\u(N,Q^2)=\int _0^1dx\,x^{N-1}\u(x,Q^2)
$$$$
\P (N,\alpha_s(Q^2))=\int_0^1 dz\,z^N\P (z,\alpha_s(Q^2))
\eqno(7)
$$
Then
$$
{\pd\over\pd t}\u (N,Q^2)= \P (N,\alpha_s(Q^2))\,\u(N,Q^2)
\eqno(8)
$$
A power contribution (1) to $F_2(x,Q^2)$ corresponds to a pole 
$$
{\f(Q^2)\over N-\epsilon_0}~~~~~~~~~~\f(Q^2)=\Big(\matrix{f_q(Q^2)\cr
f_g(Q^2)\cr}\Big )
\eqno(9)
$$
in $\u(N,Q^2)$. More generally, consider a contribution $x^{-\epsilon_0}
f(x,Q^2)$. We assume that $f(x,Q^2)$ vanishes at $x=1$ and is differentiable, 
for all $Q^2$.
Insert in the Mellin transform integral and integrate once by parts,
to get
$$
-{1\over N-\epsilon_0}\int_0^1 dx x^{N-\epsilon_0}f_x(x,Q^2)
\eqno(10)
$$
So there is a pole at $N=\epsilon_0$ with residue
$$
-\int_0^1 dx~f_x(x,Q^2)=f(0,Q^2)
\eqno(11)
$$

With 4 active quark flavours and a flavour-blind hard pomeron,
$f_q(Q^2)={18\over 5}f_0(Q^2)$.
We find\ref{\cudell}, on taking the residue of the pole at $N=\epsilon_0$
on each side of the Mellin transform (8) of the DGLAP equation,
$$
{\pd\over\pd t}\f(Q^2)= \P (N=\epsilon_0,\alpha_s(Q^2))\, \f(Q^2)
\eqno(12)
$$

We have previously\ref{\twopom} fitted accurate ZEUS and H1 data 
in the range $x<0.001$, $0.045\leq Q^2\leq 35$ GeV$^2$,
together with data for $\sigma^{\gamma p}$. The result is 
$$
\epsilon_0=0.437
\eqno(13)
$$
and
$$
Q_0^2=9.11\hbox{ GeV}^2~~~~~~~~~X_0=0.00146
\eqno(14)
$$
The error on each of these quantities is large, but we have given their
values to this accuracy because their errors are strongly correlated.

\topinsert{
\centerline{\epsfxsize=1.1\hsize\epsfbox[0 0 596 842]{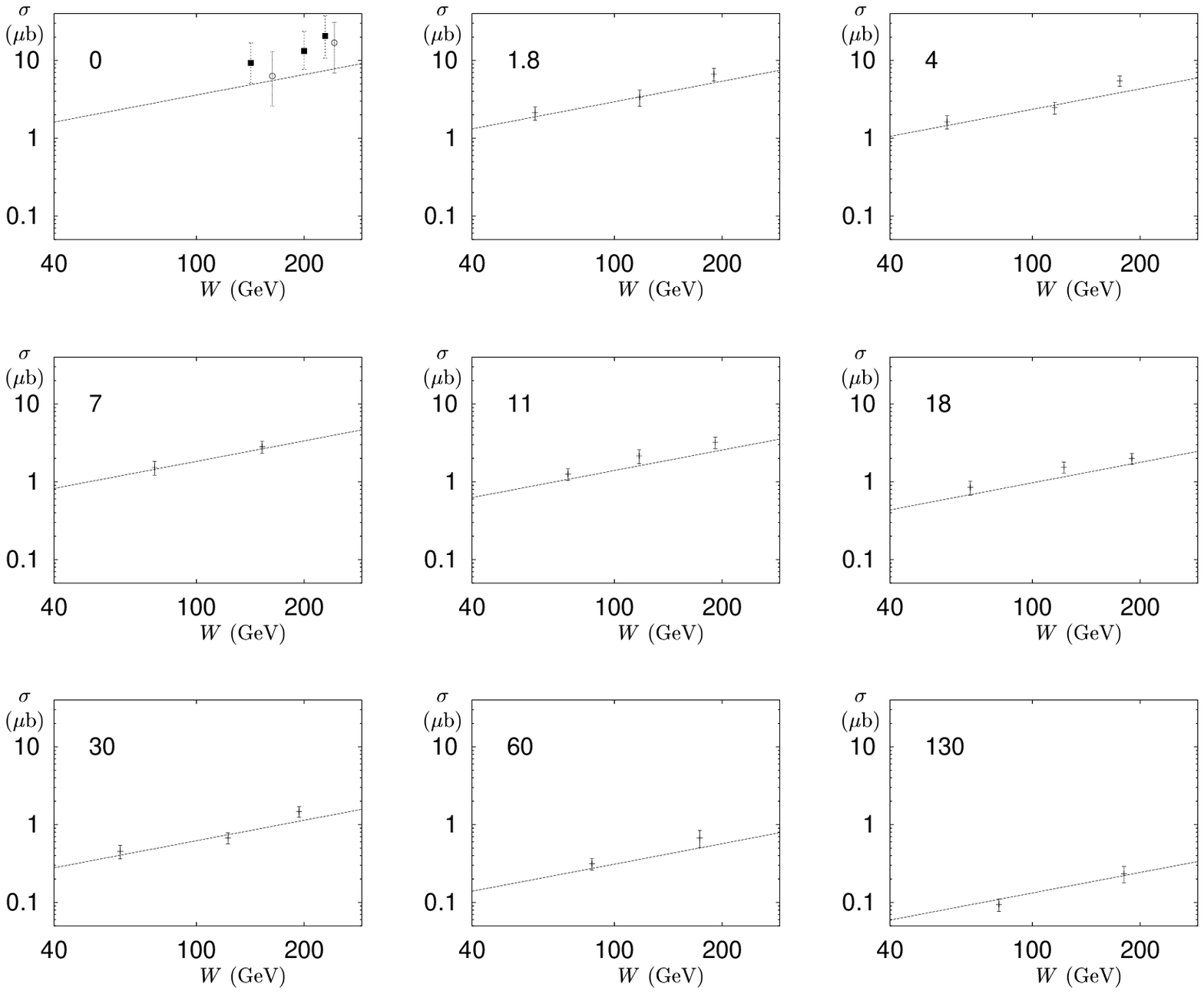}}
\vskip -12truecm
Figure 1: ZEUS data\defref\ZEUS_c{ZEUS collaboration: J Breitweg et al, 
European Physical Journal C12 (2000) 35} for the charm structure function
$F_2^c(x,Q^2)$ and the charm photoproduction cross section, with the
prediction from the fit \ref{\twopom} to $F_2(x,Q^2)$ assuming 
that the hard-pomeron is flavour blind. 
}\endinsert

If we include 4 flavours of quark and antiquark
in the sum in (6), then at $Q^2=20$ GeV$^2$ the singlet quark distribution
$x\sum _f(q_f+\bar q_f)\sim 0.095x^{-\epsilon_0}$ at sufficiently small
$x$.
We found also that the charmed-quark component $F_2^c$ of $F_2$ is
apparently governed almost entirely by hard-pomeron exchange at small
$x$, even at small values of $Q^2$,
and that, within the experimental errors,
its magnitude is consistent with the hard pomeron
being flavour-blind. The hard-pomeron description of $F_2^c$ is shown in
figure 1. 
According to perturbative QCD,
the charmed quark originates from a gluon in the proton and the two
distributions are proportional to each other to a good approximation
over a wide range of $x$ and $Q^2$\defref\SN01{Z Sullivan and P M Nadolsky, 
hep-ph/0111358}.
So this implies that the gluon
distribution also is hard-pomeron dominated. The conventional approach
to QCD evolution is correct if $x$ is not too small, because
it does not then probe the splitting matrix at $z=0$, which is where
its expansion in powers of $\alpha_s$ is illegal. We assume that at
$x=0.01$ and $Q^2=20$ GeV$^2$ the value of $g(x,Q^2)$ extracted by the HERA
experiments is reasonably close to the correct value.   
At $Q^2$=20 GeV$^2$ and
$x=0.01$, an NLO fit\defref\heragluon{
C Adloff et al, Eur Phys J C21 (2001) 33\h
A Cooper-Sarkar, in Proceedings of the EPS International Conference on
High Energy Physics, Budapest, 2001 (D Horvath, P Levai, A Patkos,
eds), JHEP (http://jhep.sissa.it/) Proceedings Section,
PrHEP-hep2001/009\h
K Nagano, www-zeus.desy.de/conferences/01/2001-Ringberg-nagano.ps.gz}
to the combined ZEUS and H1 data gives
$xg(x,Q^2)=5.7\pm 0.7$. 
Other authors\defref\mrs{
A D Martin, R G Roberts, W J Stirling and R S Thorne,
Eur Phys J C18 (2000) 117 and hep-ph/0110215
}\defref\cteq{
CTEQ Collaboration: H L Lai et al, Eur Phys J C12 (2000) 375
}\defref\durham{
Durham data base, cpt19.dur.ac.uk/hepdata/pdf3.html
} 
find much the same value. This is $8\pm 1$ times 
hard-pomeron component of the singlet quark distribution.

An unresummed perturbation expansion 
of the splitting matrix $\P(N,\alpha_s)$ 
is not valid\ref{\cudell} for small values of $N$
because it introduces a spurious singularity at $N=0$.
According to (12) and (13), we need $\P(N,\alpha_s)$ at a value of
$N$ far from 0, and so it is reasonable to hope that resummation is
not needed.
The numerical values of the elements of the matrix
$\P (N,\alpha_s)$ in one and two-loop order are plotted in figure~4.9
of the book by Ellis, Stirling and Webber\ref{\esw} 
for the value $\alpha_s=2\pi/30$.
From these we may evaluate the running-coupling splitting matrix
$\P (N,\alpha_s(Q^2))$ in one and two-loop order\footnote{$^*$}
{It is not made clear in the book that the $qg$ plot includes the
necessary factor $2n_f$.}.
We use the one and two-loop forms for the running coupling\ref{\esw} 
$$
\alpha_s^{LO}(Q^2)={1\over b t}~~~~~~~~~~~~~~~~~~~~~~~
\alpha_s^{NLO}(Q^2)={1\over b t}\Big[1-{b'\log t\over b t}\Big]
\eqno(15a)
$$
where
$$
b={33-2n_f\over 12\pi}~~~~~b'={153-19n_f\over 2\pi (33-2n_f)}
\eqno(15b)
$$
In each case, we choose $\Lambda$ such that 
$\alpha_s(M_Z^2)=0.116$. This gives
$$
\Lambda^{LO}=140 \hbox{ MeV}~~~~~~~~~~~~~~\Lambda^{NLO}=400 \hbox{ MeV}
\eqno(16)
$$
We use four flavours throughout as, at the energies we are considering,
the charm contribution is active and the beauty contribution is so small
that its omission has a negligible effect.

We integrate the differential equation (12).
The result for the singlet quark distribution is shown in figure 2a,
where the solid curve is the result of the two-loop-order
perturbative QCD evolution according to
(12), and the broken curve is the fit to the data given by (4).
We have taken the ratio of
the gluon distribution to the hard-pomeron component of the singlet quark 
distribution to be 8.0 at $Q^2=20$ GeV$^2$. Note that the gluon/quark ratio 
is a parameter which in principle we could change. However it turns out that 
the value of 8 we have obtained from the NLO fit to the combined ZEUS and H1 
data works well.
Figure 2b shows how the gluon distribution 
$$
xg(x,Q^2)=f_g(Q^2)x^{-\epsilon_0}
\eqno(17)
$$
evolves. Figure 3 shows that
there is very little difference between one-loop-order and
two-loop-order evolution, except
at small $Q^2$. This is because we have 
used the splitting function for a value of $N$ safely away from
$N=0$ and it encourages the hope that resummation, if we knew how to perform it,
would make little difference to these results.
\topinsert
\centerline{\epsfxsize=0.42\hsize\epsfbox[75 560 350 765]{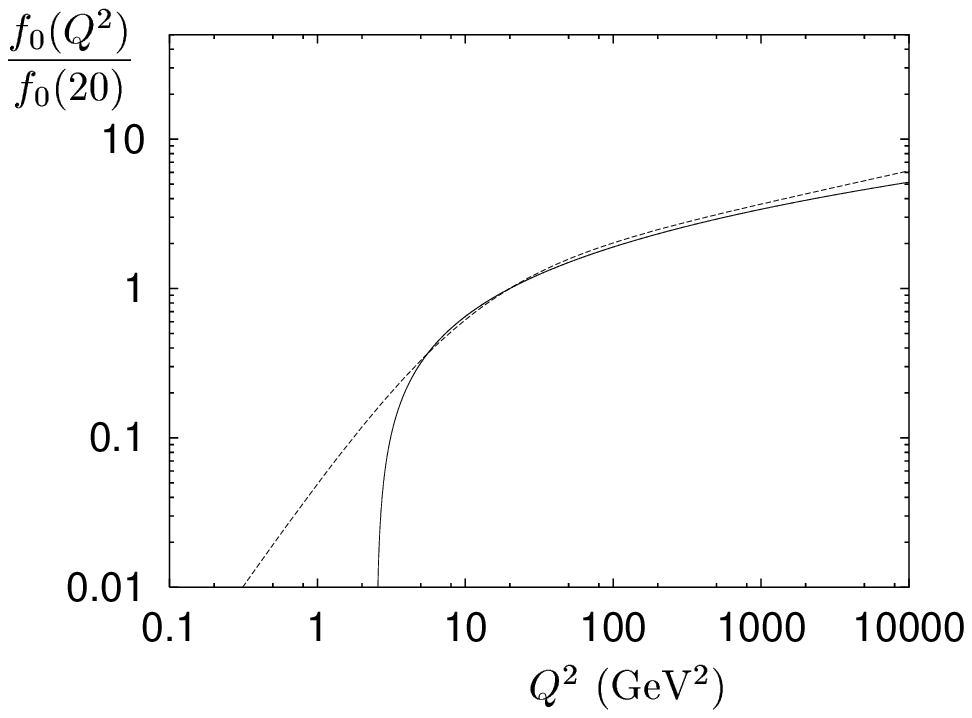}
\hfill \epsfxsize=0.47\hsize\epsfbox[60 560 345 760]{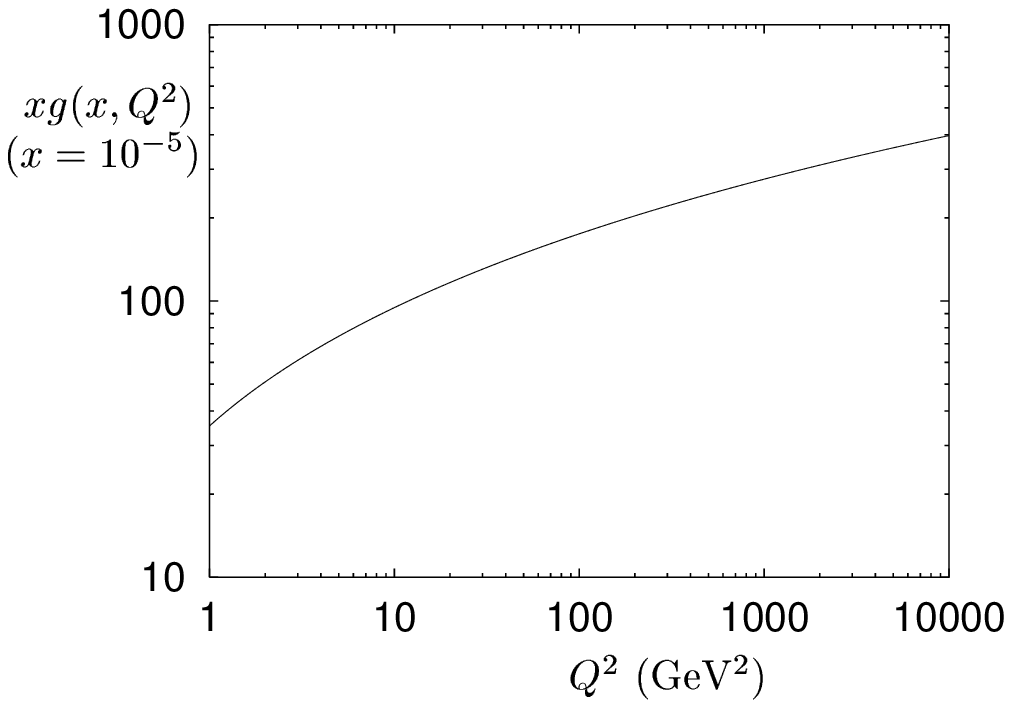}}

\line{\hfill (a)\hskip 70 truemm(b)\hfill}
Figure 2: (a) NLO evolution with $\Lambda=400$ MeV of the hard-pomeron
coefficient $f_0(Q^2)$ (solid curve) and the fit (4) (broken curve);
(b) evolution of the gluon structure function $xg(x,Q^2)$ at $x=10^{-5}$.
\vskip 7truemm
\centerline{\epsfxsize=0.42\hsize\epsfbox[70 560 340 760]{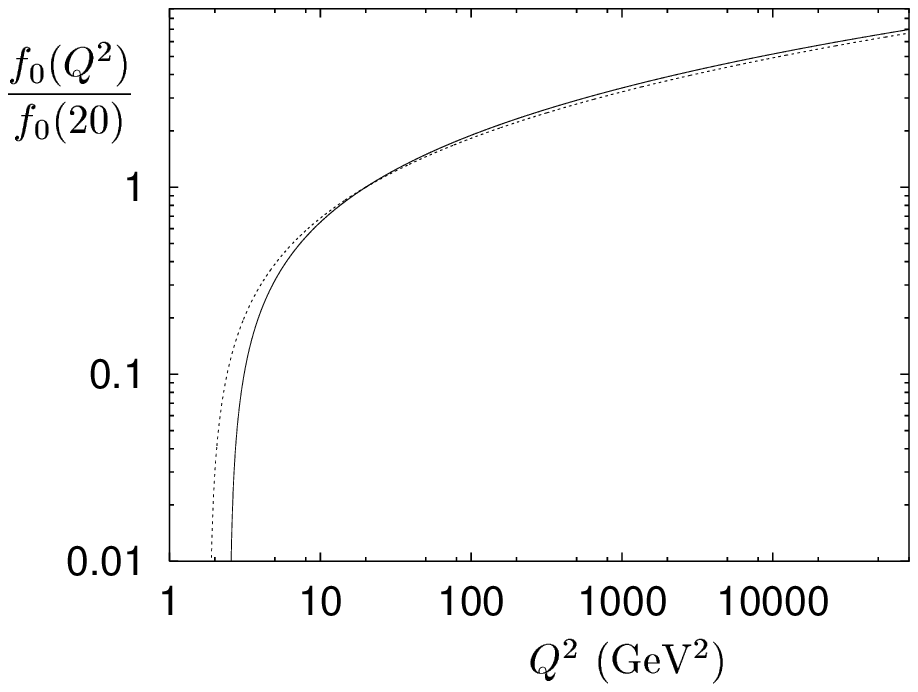}
\hfill \epsfxsize=0.444\hsize\epsfbox[60 560 345 760]{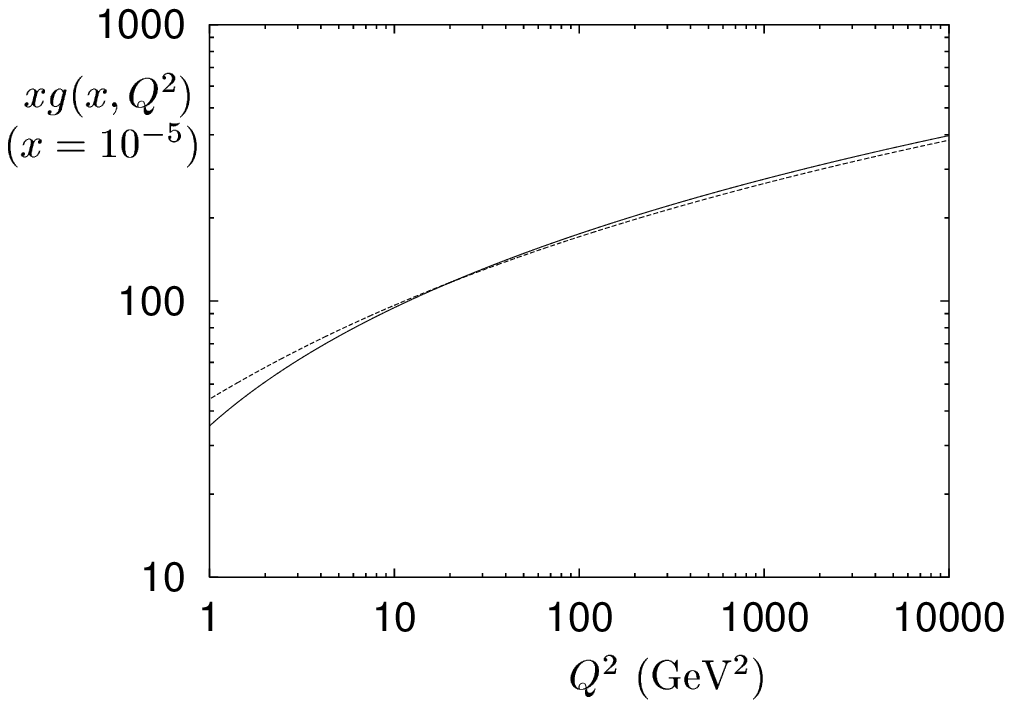}}

\line{\hfill (a)\hskip 70 truemm(b)\hfill}
Figure 3: NLO evolution with $\Lambda=400$ MeV (solid lines) and LO evolution
(broken lines) with $\Lambda=140$ MeV (a) of the hard-pomeron
coefficient $f_0(Q^2)$ and 
(b) of the evolution of the gluon structure function at $x=10^{-5}$.
\endinsert

\bigskip
{\bf 3 Discussion}

We have a number of comments on these results:

{\bf 1} It is evident from figure 2a that 
two-loop perturbative QCD describes the
evolution of the strength of the hard-pomeron contribution to $F_2(x,Q^2)$
extremely well for $Q^2$ greater than about 5 GeV$^2$, up to values
of $Q^2$ beyond where data exist. 
The perturbative QCD differential equation is a large-$Q^2$ equation and is not valid when 
$Q^2$ is small; it even makes the hard-pomeron contribution to
$F_2$ become negative below 2.5 GeV$^2$.

{\bf 2} Our fit to the data, from which we extracted the hard-pomeron
contribution, used data from $Q^2=0$ to 35 GeV$^2$. Over most of this range,
the hard-pomeron contribution is only a small part of
$F_2(x,Q^2)$, though its relative magnitude has increased by 
$Q^2=35$ GeV$^2$: see figure 4.
So, although it is well-established\ref{\mrs,\cteq} that the complete
$F_2(x,Q^2)$
obeys perturbative QCD evolution beyond $Q^2=5$ GeV$^2$ for the range of $x$ where
data exist, it is not at all trivial
that the hard-pomeron part of it does. This successful link with
perturbative QCD provides rather striking
verification of the hard-pomeron concept and of our having extracted
it correctly from the data. 

{\bf 3} It also provides a clean
test of perturbative QCD itself, 
one that genuinely tests evolution rather than being a 
global fit\ref{\mrs,\cteq}.
It depends on just one parameter, the ratio of the gluon distribution
to the hard-pomeron part of the singlet quark distribution
at some value of $Q^2$, which we chose to be 20 GeV$^2$. The curve in figure~%
2a uses the value 8.0 for this ratio but, as we have said, analysis
of the HERA data gives an error $\pm 1$ on this number. Changing it within
this range has a fairly small effect at large $Q^2$, about 10\% at
5000 GeV$^2$, but causes quite
a large change in the details of how it breaks away from the 
phenomenological curve when we evolve back to
small $Q^2$. Similar remarks apply to changing
$\Lambda$ within its allowed range of values.
\topinsert{
\centerline{\epsfxsize=0.5\hsize\epsfbox[80 610 290 760]{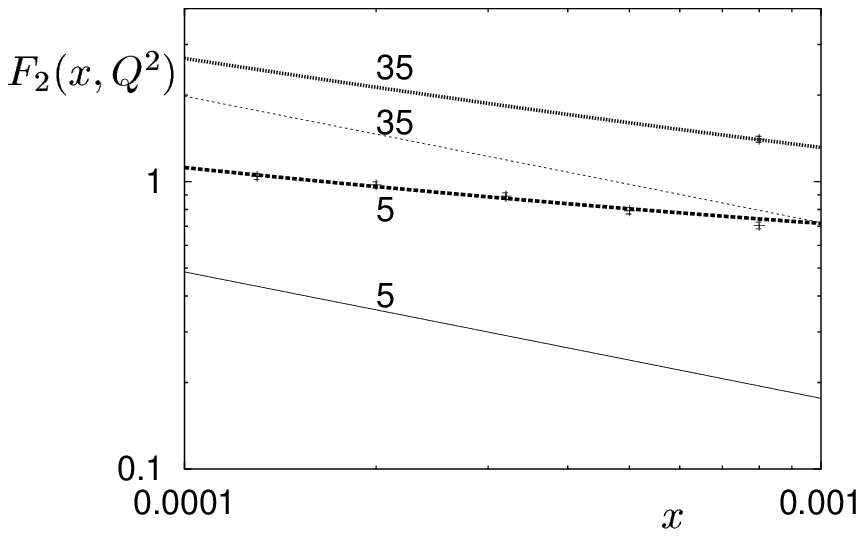}}

Figure 4: The thick curves are fits to $F_2(x,Q^2)$ at $Q^2=5$ and 
$35$ GeV$^2$, with the H1 data used at those values of $Q^2$.
The thin curves are the hard-pomeron contributions.
}\endinsert

{\bf 4} Our phenomenological fit behaves
at large $Q^2$ as the power $Q^{\epsilon_0}$, while the
solution to the perturbative QCD differential equation (12) 
rather behaves asymptotically
as a power of $\log Q^2$, approximately $(\log Q^2)^{2}$; 
nevertheless the two agree well over a very large range of $Q^2$. 
According to figure 2a, they begin to diverge from each other
only at values of $Q^2$ beyond the data, so in our fit we
could instead use the perturbative QCD form. However, it is
very much more complicated to parametrise and, more importantly,
using the perturbative QCD form at small $Q^2$ is incorrect. 
Using different
forms for different $Q^2$ ranges would lead to matching problems and
it would be almost impossible to achieve the required analyticity in
$Q^2$ at the join. 

{\bf 5} The conventional approach to evolution expands the splitting
matrix $\P(N,\alpha_s)$  in powers of $\alpha_s$. Successive terms in
the expansion are increasingly singular at $N=0$. This is a signal that
the expansion is illegal\ref{\cudell} for small values of $N$, 
since the complete splitting matrix is regular at $N=0$.
Because an unresummed expansion that needs the splitting matrix
at small $N$  makes the splitting function
larger than it really is, a gluon distribution of a given
magnitude apparently gives stronger evolution than it really
should. That is, the conventional approach will tend to
under-estimate the magnitude of $xg(x,Q^2)$ in certain regions
of $(x,Q^2)$ space. This is verified by our results for the evolution of
$xg(x,Q^2)$: figure 5 shows the proton's gluon structure function at 
two values of $Q^2$, according to our calculations, which do not use
the splitting matrix at small $N$, and compares it
with what is extracted from the data by conventional means.
\topinsert{
\centerline{\epsfxsize=0.45\hsize\epsfbox[50 560 330 760]{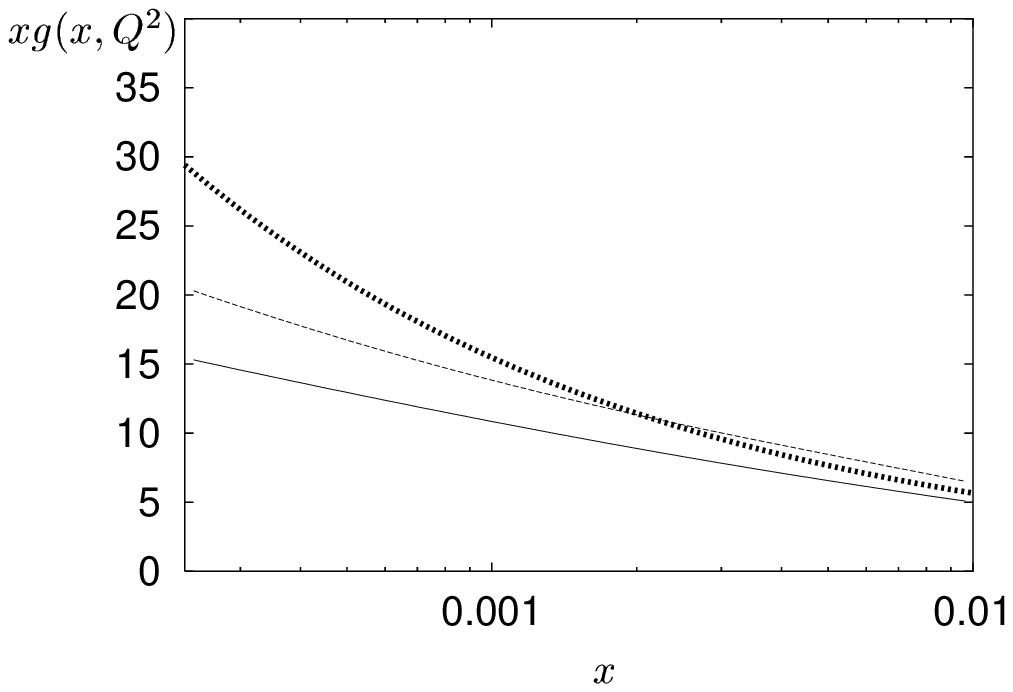}
\hfill
\epsfxsize=0.45\hsize\epsfbox[60 565 345 755]{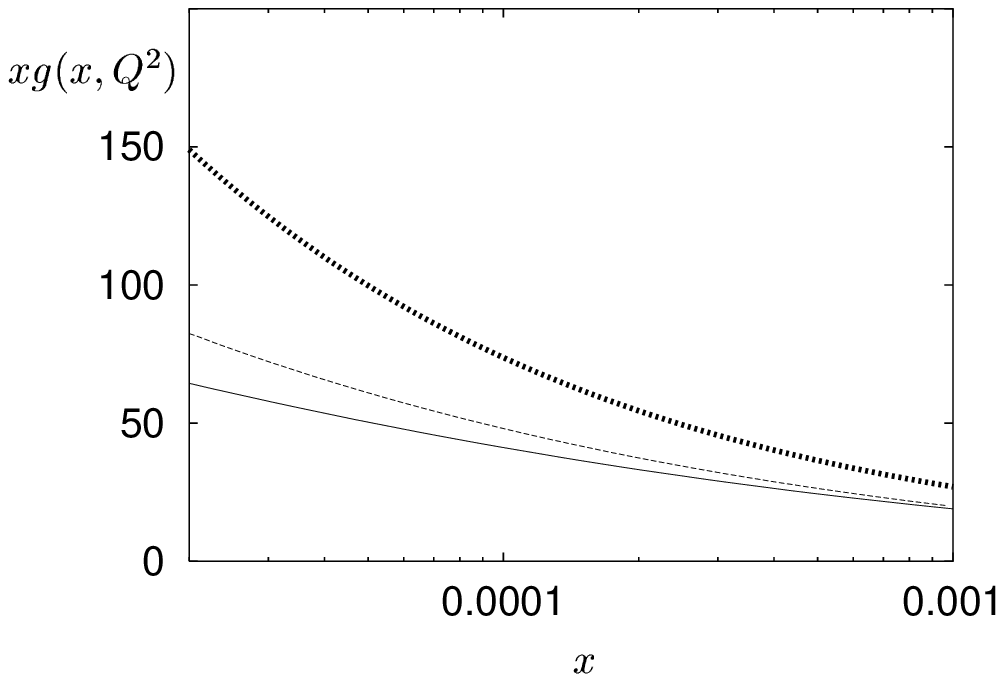}}

\line{\hfill (a)\hskip 70 truemm(b)\hfill}
Figure 5: Gluon structure function $xg(x,Q^2)$ at (a) $Q^2=20$ and
(b) $200$ GeV$^2$.
In each case the thick line is our evolved distribution. In (a) the thin
lines are the limits extracted by conventional NLO analysis of HERA 
data\ref{\heragluon}. In (b) the middle line is\ref{\cteq,\durham} CTEQ5M and the 
lower line is\ref{\mrs,\durham} MRST20011.}
\endinsert

{\bf 6} We cannot use a similar approach to the evolution of the
contribution from the soft pomeron, because this needs $\P (N,\alpha_s)$
at $N\approx 0.08$, dangerously close to $N=0$. Handling this will need
resummation\defref\ciafaloni{
M Ciafaloni, D Colferai and G P Salam,
JHEP 0007 (2000) 054
}\ref{\thorne}\ref{\forte},
so as to tame the singularity at $N=0$, but we do not yet have a reliable
way to perform this resummation. As we have explained, experiment finds
that the charm-quark distribution does not 
contain a soft-pomeron term; we are assuming that 
our inability to handle the soft pomeron 
is therefore not relevant for the gluon distribution.

{\bf 7} The fact that we can fit the data so well with the only singularities
of $\u (N,Q^2)$ in $N\geq 0$ identified with the hard and soft pomerons,
supports our belief\ref{\cudell} that $\P (N,\alpha_s)$ does
not have any singularities in this region. This again contrasts with the
conventional attitude: we maintain that
the relevant singularities of  $\u (N,Q^2)$ in the complex $N$-plane are
not {\it generated} by the perturbative evolution, which merely governs
how the strength of their contribution increases with $Q^2$. They
are already present at $Q^2=0$ and their position does not vary with
$Q^2$. We have explained before\ref{\cudell} that there are strong theoretical
reasons to believe this: one cannot simply start perturbative QCD evolution at some value
of $Q^2$ and ignore how this joins on to what is found for smaller values.

{\bf 8} We stress again how very few parameters we need to fit the
data for $F_2(x,Q^2)$ for $x<0.001$. 
The hard-pomeron contribution is parametrised in terms of the three
parameters $X_0,\epsilon_0$ and $Q_0$ introduced in (4),
and for the soft pomeron we have similarly\ref{\twopom}
$X_1,\epsilon_1$ and $Q_1$:
$$
f_1(Q^2)=X_1~(Q^2)^{1+\epsilon_1}/(1+Q^2/Q_1^2)^{1+\epsilon_1}
\eqno(18)
$$
Of these, $\epsilon_1\approx 0.08$ is determined from soft hadronic 
interactions, and $X_1$ from the fit to $\sigma^{\gamma p}$. The latter
requires\ref{\twopom} also a contribution from $f_2$ and $a_2$ exchange, 
but this does not contribute very much to $F_2(x,Q^2)$ for $x<0.001$.

{\bf 9} We have remarked that, although the data for $F_2(x,Q^2)$ are now
highly accurate, the value of $\epsilon_0$ is still quite poorly
determined. It depends\ref{\twopom} on what is assumed  for the
large-$Q^2$ behaviour of the soft-pomeron coefficient function
$f_1(Q^2)$. In this paper we have used $\epsilon_0=0.437$, which
corresponds to asuming that $f_1(Q^2)$ has the form (18)
and so goes to a constant at large $Q^2$. But one may obtain an equally
good fit to the data for $F_2(x,Q^2)$ by assuming that $f_1(Q^2)$
vanishes like $1/Q$ at large $Q^2$.
Figure~6a shows the equivalent of figure~2a
for this case, with $\epsilon_0=0.394$, 
and figure~6b is the equivalent of figure~5a, from which it
is seen that the gluon structure function needs to be slightly
larger at $Q^2=20$ GeV$^2$ than in the previous case. By $Q^2=2000$ GeV$^2$
it is quite a lot larger: see figure 7. The lower curve in this figure
is the CTEQ4M prediction, based on conventional unresummed NLO 
evolution{\cteq}.
\topinsert{
\centerline{\epsfxsize=0.45\hsize\epsfbox[50 560 330 760]{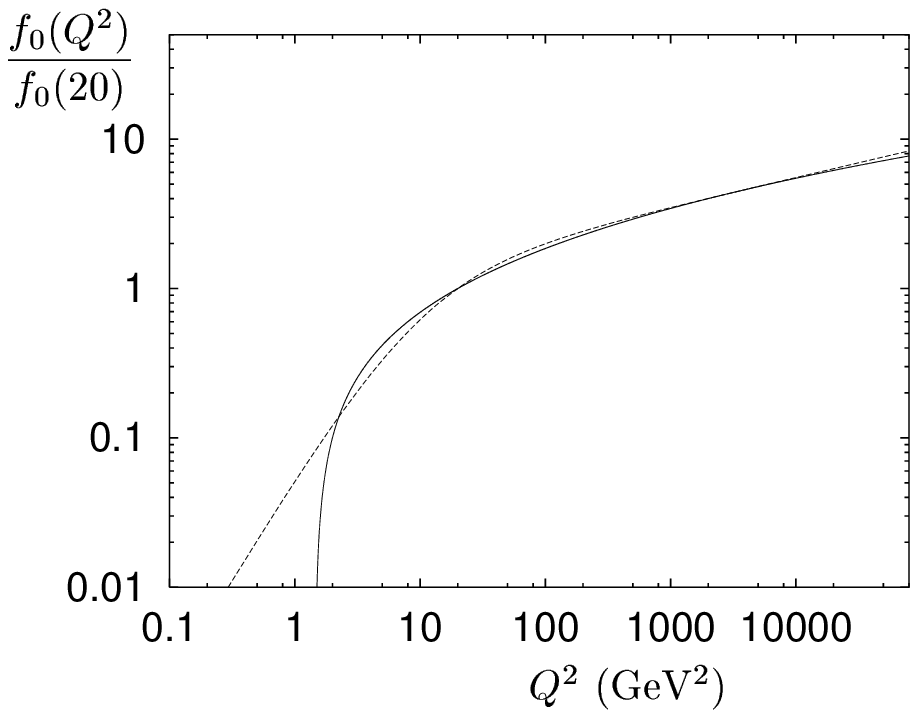}
\hfill
\epsfxsize=0.45\hsize\epsfbox[50 560 330 760]{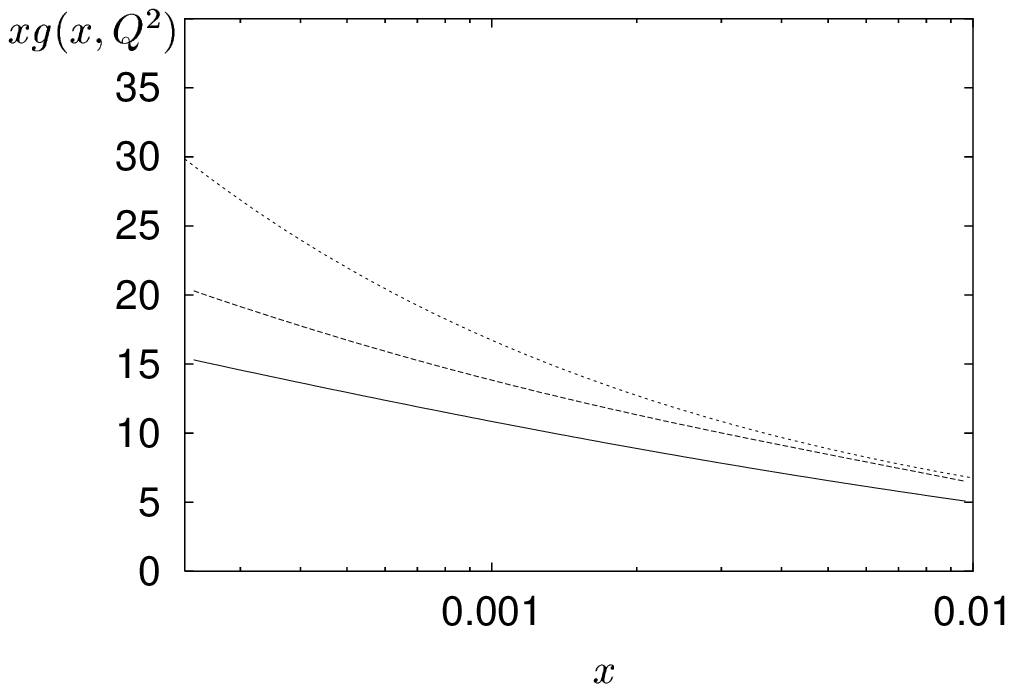}}

\line{\hfill (a)\hskip 70 truemm(b)\hfill}
Figure 6: Equivalents of (a) figure~2a  and (b) figure~5a for
the case $\epsilon_0=0.394$
\vskip 7truemm
\centerline{\epsfxsize=0.45\hsize\epsfbox[55 590 325 755]{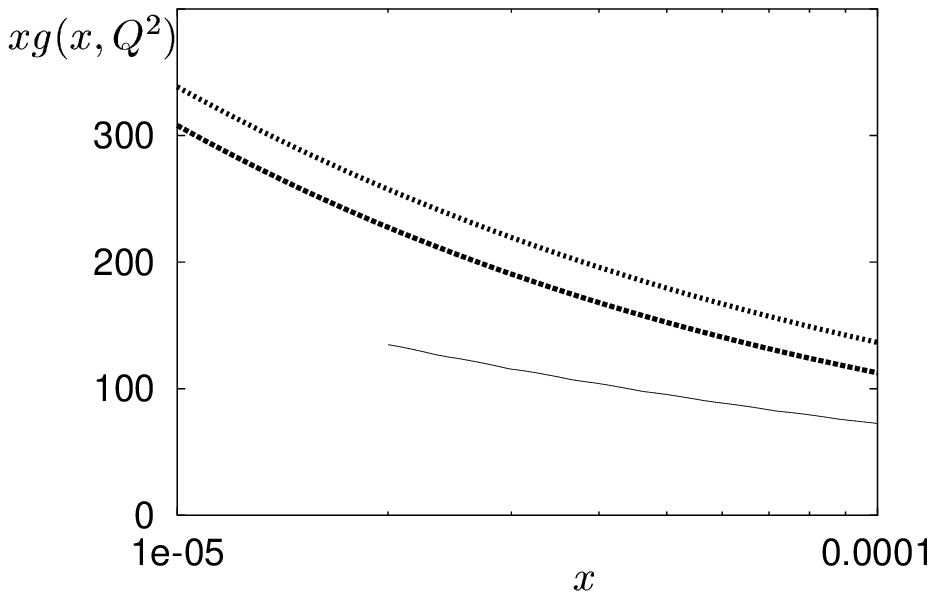}}
\vskip -1truemm
Figure 7: Gluon distribution at $Q^2=2000$ GeV$^2$. 
The upper line is the perturbative QCD-evolved distribution for $\epsilon_0=0.394$,
the middle line for $\epsilon_0=0.437$, and the lower line is 
the MRST20011 two-loop prediction\ref{\mrs,\durham}
}\endinsert

{\bf 10} The LO or NLO
perturbative QCD evolution makes the coefficient $f_0(Q^2)$ of the leading
power of $1/x$ increase indefinitely with increasing $Q^2$. In order
not to conflict with the momentum sum rule, at any fixed $x$
nonleading powers of $x$ must become progressively more important, so
that the largest value of $x$ for which the leading power alone gives a
good approximation to $F_2(x,Q^2)$ decreases as $Q^2$ 
increases\footnote{$^{\dag}$}{We are grateful to Otto Nachtmann, Robert Thorne
and our referee
for emphasising this to us}. This is why, while our plots of the gluon 
distribution at $Q^2=20$ GeV$^2$ extend up to $x=0.01$, for 200 GeV$^2$
we stop at x=0.001, and for 2000 GeV$^2$ at 0.0001. We guess that
these values for the
limits of the validity of the single-term approximation are safe.

{\bf 11} To summarise, what we have achieved is a genuine evolution, in contrast
to global fits\ref{\mrs,\cteq} which use data at large $Q^2$ to help
constrain the fit already at small $Q^2$. We used data up to $Q^2=35$~GeV$^2$
to determine the hard-pomeron component of $F_2(x,Q^2)$; in this fit we did
not assume perturbative QCD evolution, but just made a 
numerical fit to the data\ref{\twopom}.
Now we find that, for the hard-pomeron contribution, we have almost perfect agreement
with perturbative QCD evolution up to $Q^2=5000$~GeV$^2$. This is a striking verification
that the hard-pomeron concept is correct, as well as being a  
success for perturbative QCD itself. It leads us to conclude that, 
at very small $x$,
the gluon structure function is somewhat larger than
has until now been believed.
\bigskip{\eightit
This research is supported in part by the EU Programme
``Training and Mobility of Researchers", Network
``Quantum Chromodynamics and the Deep Structure of
Elementary Particles'' (contract FMRX-CT98-0194),
and by PPARC}
\vfill\eject
{\medskip\immediate\closeout\rfile\writestoppt
\baselineskip=7pt{{\bf References}}\bigskip{\frenchspacing%
\parindent=20pt\escapechar=` \input refs.tmp\bigskip}\nonfrenchspacing}
\bye